\documentclass{PoS}

\title{Limits for primordial magnetic fields (invited talk)}

\ShortTitle{Limits for primordial magnetic fields (invited talk)}

\author{\speaker{Chiara Caprini}
        \thanks{I wish to thank Ruth Durrer and Andrea Ferrara, the organisers of the parallel session "Intergalactic magnetic and radiation fields", for the invitation.}\\
  CNRS, URA 2306  \& CEA, IPhT, F-91191 Gif-sur-Yvette, France\\
       E-mail: \email{chiara.caprini@cea.fr}}

\abstract{A possible explanation for the origin of the magnetic fields observed today in matter structures is that they were generated in the primordial universe. After briefly revising the model of a primordial stochastic magnetic field and sketching the main features of its time evolution in the primordial plasma, we illustrate the current upper bounds on the magnetic field amplitude and spectral index from Cosmic Microwave Background observations and gravitational wave production. We conclude that a primordial magnetic field generated by a non-causal process such as inflation with a red spectrum seems to be favoured as a seed for the magnetic fields observed today in structures.}

\FullConference{25th Texas Symposium on Relativistic Astrophysics \\
		 December 6-10, 2010\\
		 Heidelberg, Germany}

\newcommand{\ie}{{\em i.e. }} 
\newcommand{\eg}{{\em e.g. }} 
\newcommand{\cf}{\emph{cf.~}}

\newcommand{\lag}{\langle} 
\newcommand{\rag}{\rangle}

\newcommand{\Om}{\Omega}

\newcommand{\be}{\begin{equation}} 
\newcommand{\ee}{\end{equation}} 
\newcommand{\bee}{\begin{equation}} 
\newcommand{\eee}{\end{equation}} 
\newcommand{\bea}{\begin{eqnarray}} 
\newcommand{\eea}{\end{eqnarray}} 
\newcommand{\bean}{\begin{eqnarray*}} 
\newcommand{\eean}{\end{eqnarray*}}

\newcommand{\bB}{{\mathbf B}} 
\newcommand{\bx}{{\mathbf x}}

\begin{document}

\section{Introduction}

The observational evidence of large scale magnetic fields in matter structures covers nowadays an impressive range of length-scales and redshift: from galaxies to regions around high redshift quasars, from clusters and superclusters to low density filamentary regions \cite{filaments}. Recently, Fermi and HESS data have been used to put a lower bound on the intensity of the magnetic field in the inter-galactic medium: $B \gtrsim 10^{-15}$ G \cite{inter}. Observations show that magnetic fields of $\mu$G intensity are universally present in galaxies and clusters, and they are correlated on scales of the order of the galaxy or cluster size; they must have grown in a relatively short time, since they are observed also at high redshift; they seem to be present also in filaments and voids although presumably with lower intensity. These characteristics make it extremely difficult to find an explanation for their origin, which remains to date an open problem. One of the possible explanations is that they have been generated in the primordial universe, and this is the case treated in this paper. Magnetogenesis mechanisms operating in the early universe have the advantage to provide magnetic seeds filling the entire universe: this goes in the right direction to explain both the ubiquity of the observed fields and the uniformity of the measured amplitudes. The initial magnetic seed is amplified during the process of structure formation: to explain the $\mu$G fields observed today in clusters, a seed of the order of the nG (redshifted to today) is probably enough \cite{grasso}. In the case of galaxies, the mean field dynamo can amplify an initial seed of the order of $10^{-23}$ G (redshifted to today) to $\mu$G level \cite{grasso}. It is usually required that these primordial seed amplitudes be present on a scale of about $\lambda \sim 100$ kpc, corresponding to the minimal scale that survives damping by photon viscosity, as given in \cite{subrabarrow}\footnote{According to \cite{jedam}, helical primordial fields can survive damping by photon viscosity on smaller scales, of the order of $10$ kpc, and in principle this could also be sufficient to seed the observed fields. In general, assessing at which scale $\lambda$ the magnetic field should be present before structure formation in order to explain the observations is a complicated matter, which can be addressed only with numerical simulations \cite{dolag}.}. There have been many proposals in the literature up to date on how to generate the seeds for the observed magnetic fields. Primordial generation mechanisms operating before the epoch of recombination can be divided in two classes (for a review, see \cite{kandus}). Causal mechanisms take place when the universe has a finite causal horizon; they are based either on charge separation arising at the bubble walls during a first order electroweak (EW) or QCD phase transition, or on the properties of the Lagrangian of the EW interaction. Non-causal mechanisms like inflation operate when the causal horizon of the universe diverges; they are based on non-standard Lagrangians which break the conformal invariance of electromagnetism. Both classes give rise to a stochastic magnetic field, statistically homogeneous and isotropic: in section \ref{sec:model} we review how to model such a magnetic field and how it evolves in time. If a primordial magnetic field is present in the early universe, it is possible to use observables such as the Cosmic Microwave Background (CMB) or Nucleosynthesis to infer limits on its intensity: in sections \ref{sec:cmb} and \ref{sec:nuc} we review the current observational and theoretical constraints that can be put on the magnetic field amplitude in correlation with its generation mechanism, both using the CMB and gravitational waves (GWs) at Nucleosynthesis. 

\section{Magnetic field model and time evolution}
\label{sec:model}

The power spectrum of a primordial stochastic magnetic field, statistically homogeneous and isotropic, is given by two terms, one representing the magnetic field energy density, the other representing the helicity density:
\begin{eqnarray} 
& & \langle B_i({\mathbf k})B^{*}_j({\mathbf q})\rangle 
=\frac{(2\pi)^3}{2} \delta({\mathbf k}-{\mathbf q}) [(\delta_{ij}-\hat{k}_i\hat{k}_j) \, S(k)   + 
i \epsilon_{ijm} \hat{k}_m \, A(k)]\,, 
\label{spectrum}  \\
& &\rho_B=\frac{1}{2}\lag \bB(\bx)\cdot\bB(\bx)\rag=\frac{1}{(2\pi)^2}\int_0^\infty
dk\, k^2\, S(k)\,, ~~~
H=\lag {\bf A}(\bx)\cdot \bB(\bx)\rag=\frac{2}{(2\pi)^2}\int_0^\infty
dk\, k\, A(k)\,, \nonumber
\end{eqnarray} 
where ${\bf A}(\bx)$ denotes the vector potential. The helicity is a measure of the links and twists of the magnetic field lines; a magnetic field is said to be maximally helical when the condition $S(k)=|A(k)|$ is satisfied \cite{dynamo}. The helical part of the spectrum does not contribute to the energy density, but it influences the time evolution of the magnetic fields, as we will see. At sufficiently large scales, the magnetic field power spectrum is not expected to have any structure, and can be approximated by a simple power law: $S(k\rightarrow 0)\propto k^n$. In order for the energy density to remain finite, one has then $n>-3$. This power law behaviour is in general maintained up to the wavenumber corresponding to the correlation scale $k_L=2\pi /L$. The correlation scale $L$ depends on the process that generated the magnetic field: for a causal generation mechanism, by definition it must be smaller or equal to the cosmological horizon at the moment of generation $L\leq \eta_*$, where $\eta$ denotes conformal time. Causality also puts a constraint on the spectral index at large scales: since the correlation function of a causal magnetic field has compact support, \ie it decays to zero on scales larger than the correlation scale, its Fourier transform (the power spectrum) must be analytic. From Eq.~\ref{spectrum} one sees that analyticity of the power spectrum implies $n\geq 2$ and an even integer \cite{causa}\footnote{Although it seems that the time evolution can modify the spectral index on scales $\eta_*^{-1}<k<L^{-1}$~\cite{jedamsigl}.}.

Before neutrino decoupling and electron-positron annihilation ($T\gtrsim 1$ MeV), the viscosity of the primordial plasma is low: the universe is in a turbulent state with very high Reynolds number. Therefore, at scales smaller than the correlation scale $L$ the magnetic power spectrum develops a turbulent tail and decays as a power law that could be, for example, of the Kolmogorov type $k^{-11/3}$. The power law decay extends up to the damping scale $k_D$, determined by the kinetic viscosity: at smaller scales, the Reynolds number becomes smaller than one and the magnetic energy is dissipated. $k_D$ is therefore the upper cutoff of the magnetic power spectrum. Most importantly, the magnetohydrodynamic (MHD) turbulent cascade influences also the time evolution of the magnetic field, beyond the decay due to the expansion of the universe $B\propto a^{-2}(\eta)$ (where $a(\eta)$ is the scale factor). The free decay of the MHD turbulence has been studied theoretically, experimentally and by numerical simulations \cite{theor,jedam}. The resulting time evolution of the magnetic spectrum is such that both the correlation and the damping scales grow in time as power laws, while the magnetic energy density decays, also as a power law. There is no consensus on the actual value of the power law exponents, but both analytical analyses and numerical simulations so far do agree on one point: while a non-helical magnetic field evolves following a direct cascade, in which the large scale part of the magnetic spectrum is constant in time, a helical field undergoes an inverse cascade, meaning that the magnetic energy is transferred to larger and larger scales \cite{theor,elisa}. This kind of evolution can be very important for primordial magnetic fields: the fact that the magnetic energy is dissipated at a slow rate, and that it can be transferred to large, probably even galactic scales, goes in the right direction to favour the primordial origin of the presently observed fields. In the following, we assume that the time dependence of the magnetic field energy is $\rho_B(\eta) L^{n+3} (\eta) =$ constant for the direct cascade, while $\rho_B(\eta) L (\eta) =$ constant for the inverse cascade; the time dependence of the dissipation scale $k_D(\eta)$ follows, once the kinetic viscosity is known (for details, see \cite{elisa}; note that all quantities are comoving - we eliminate  the scaling with redshift). The free decay of the MHD turbulent cascade stops when the entire turbulent range of the magnetic field spectrum is dissipated, \ie when $L(\eta_{\rm fin})\simeq 1/k_D(\eta_{\rm fin})$ \cite{elisa}. Afterwards, for $T\lesssim 1$ MeV, the kinetic viscosity of the plasma has a sudden increase, and the universe enters a viscous phase: the time evolution of the magnetic field is ruled by the formation of MHD waves in a viscous plasma (for details, see \cite{subrabarrow,jedam}). 

The (comoving) magnetic field energy density parameter is defined by $\Omega_B=\rho_B/\rho_c$, where $\rho_c$ is the critical energy density today. Related to this, is the variance of the magnetic field amplitude on a given scale $\lambda$: 
\be
B_\lambda = \frac{1}{2\pi^2} \int_0^{k_D} dk\,k^2\, S(k) \, {\rm e}^{ -\frac{k^2\lambda^2}{2} }\,,~~~~~~~
\Om_B\propto \frac{B_\lambda^2}{\rho_c} (\lambda k_D)^{n+3}\,, ~~~~~~~
 \frac{\Om_B}{\Om_{\rm rad}}(\eta_0)  \simeq 10^{-7}\frac{\langle B^2(\eta_0)\rangle}{(10^{-9}\,{\rm G})^2}\,.
\label{OmB}
\ee
To derive the second equality we have neglected the turbulent tail of the spectrum and identified $L\simeq 1/k_D$ \cite{elisa}. $\Omega_B(\eta)$ depends on time solely via the dissipation of the magnetic energy due to the MHD cascade or to viscous effects \cite{elisa}. The third equality is obtained by a straightforward numerical estimate, with $\rho_{\rm rad}(\eta_0) \simeq 2 \cdot 10^{-51} {\rm GeV}^4$; $\eta_0$ denotes comoving time today.

\section{Constraints on a primordial magnetic field from the CMB}
\label{sec:cmb}

The most convincing way to establish whether a magnetic field, capable to seed those observed today in the structures, has been generated in the primordial universe, would be to detect its trace in the CMB. The phenomenology of the imprint of a primordial magnetic field in the CMB is extremely rich; to date there is no detection, but upper bounds on the field intensity of the order of the nG have been established using CMB data. A primordial magnetic field induces scalar, vector and tensor perturbations in the metric, leading to both temperature anisotropies and polarisation signals. Analytical estimates of the CMB temperature spectrum at low multipoles $\ell \lesssim 60$ give a scaling with the magnetic field parameters as $ \ell^2 C_\ell \propto \left( \Om_B/\Om_{\rm rad} \right)^2 \left[ \ell / (k_D(\eta_{\rm rec}) \eta_0) \right]^{f(n)}$ \cite{helical,nong}. The value of the damping scale at recombination is such that $k_D(\eta_{\rm rec}) \eta_0 \gg 1$, and the spectral dependence $f(n)$ is $f(n)=2n+6$ for $n<-3/2$, and always a positive power for $n>-3/2$ : from Eq.~\ref{OmB}, we can therefore expect a constraint of the order of a nG for red spectral indexes $n\rightarrow -3$. This is indeed what found in more refined numerical analyses. 

The most relevant contribution to the CMB turns out to be from the vector modes due to Alfv\'{e}n waves, which do not suffer Silk damping and can therefore overcome, at high multipoles, both the primary temperature spectrum and the B polarisation signal due to lensing \cite{alfven}. There have been several numerical and analytical studies of the CMB spectra due to a primordial magnetic field, \eg \cite{cmb,finelli,shaw}. To date, CMB data only (WMAP7 + QUAD + ACBAR, marginalised over the other cosmological parameters) constrain the magnetic field amplitude smoothed over a scale of 1 Mpc to $B_{1\,{\rm Mpc}}< 5 $ nG, and favour a red magnetic field spectral index $n<-0.12$ \cite{finelli}. Since the magnetic field effect is mainly at small scales, it also changes the shape of the matter power spectrum: consequently, the constraints improve including the SZ effect from South Pole Telescope data and Lyman $\alpha$ data from SDSS, constraining the amplitude down to $B_{1\,{\rm Mpc}}< 1.3$ nG \cite{shaw}. 

Other interesting effects of a primordial magnetic field on the CMB are non-gaussianities \cite{nongvari,nong}, Faraday rotation of the primordial polarization \cite{faraday}, and parity-odd cross correlations (TB and EB) \cite{helical}. These latter are identically zero in a parity invariant universe, and would be a very distinctive signal of a helical magnetic field. However, the constraints on the magnetic field amplitude derived up to now from these processes are not particularly significant compared to those discussed above.

\section{Constraints from gravitational waves at Nucleosynthesis}
\label{sec:nuc}

Any extra radiation-like component present in the universe prior to Nucleosynthesis is constrained to be $\Om_{\rm rel}\leq 0.1 \, \Om_{\rm rad}$. This can be used to infer an upper bound on the magnetic field amplitude: since the magnetic field is radiation-like, naively one would apply the bound directly to the magnetic field energy density at Nucleosynthesis time $\Om_B(\eta_{\rm nuc})$. However, more stringent constraints arise considering that a primordial magnetic field, since the moment of its formation $\eta_*$, sources GWs through its anisotropic stresses \cite{elisa,gwold}. Once generated, GWs propagate freely in the universe, while the magnetic field energy is subject to dissipation due to the turbulent MHD cascade and to plasma viscosity. Therefore, part of the magnetic field energy can be "stored" into GW energy before being dissipated into heat. Denoting $\Om_B^*$ the magnetic field energy density at generation time (\eg inflation, the EW or QCD phase transitions), the GW energy density today can be written as $\Om_{\rm GW}=\mathcal{E} (\Om_B^*)^2/\Om_{\rm rad}$, where $\Om_{\rm rad}$ denotes the radiation energy density today, and $\mathcal{E}$ is an efficiency factor that can be quite big, $\mathcal{E} \lesssim 1$: a sizable fraction of the magnetic field energy is converted to GWs before dissipation \cite{gwold}. Setting $\Om_{\rm GW}\leq 0.1 \, \Om_{\rm rad}$ one therefore obtains $\Om_B^*\leq \sqrt{0.1/\mathcal{E}}\,\Om_{\rm rad}$, while setting $\Om_B(\eta_{\rm nuc})\leq 0.1 \, \Om_{\rm rad}$, the constraint on $\Om_B^*$ is weakened by a factor $\sqrt{0.1\mathcal{E}} \,\, [\Om_B^*/\Om_B(\eta_{\rm nuc})] =  \sqrt{0.1\mathcal{E}}\,\,[k_D^*/k_D(\eta_{\rm nuc})]^{n+3}\,\,[B_\lambda^*/B_\lambda(\eta_{\rm nuc})]^2 \gg 1$. For the first equality we have used Eq.~\ref{OmB}; the second inequality comes from the growth in time of the damping scale and the decay of the magnetic energy; moreover, $n>-3$. Depending on the details of the magnetic field evolution in time, on its spectral index and on the efficiency $\mathcal{E}$, this factor can become of several orders of magnitude \cite{gwold} \footnote{In principle the Nucleosynthesis bound should be applied to the total radiation-like energy density coming from the magnetic field, \ie not only $\Om_{\rm GW}\leq 0.1 \, \Om_{\rm rad}$ but $\Om_{\rm GW}+\Om_B(\eta_{\rm nuc})\leq 0.1 \, \Om_{\rm rad}$, where the part which has already been dissipated into heat by Nucleosynthesis time must be included in $\Om_{\rm rad}$, the total radiation energy density today (which is a measured quantity). However, $\Om_B(\eta_{\rm nuc})$ can be neglected with respect to $\Om_{\rm GW}$, as shown above.}.

\begin{figure} 
\includegraphics[width=.5\textwidth]{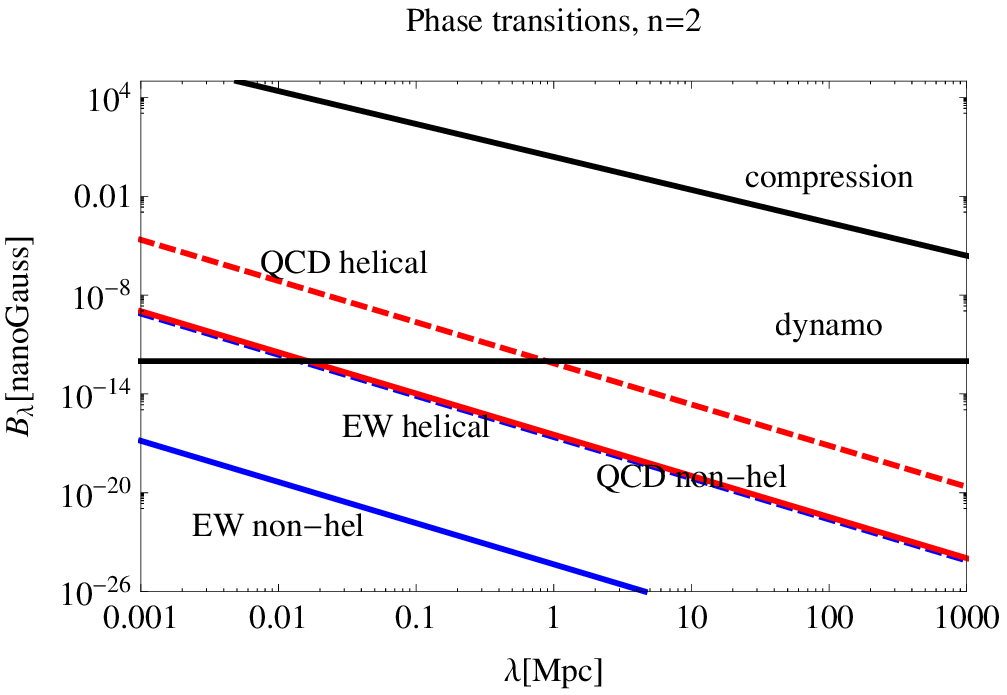} 
\includegraphics[width=.5\textwidth]{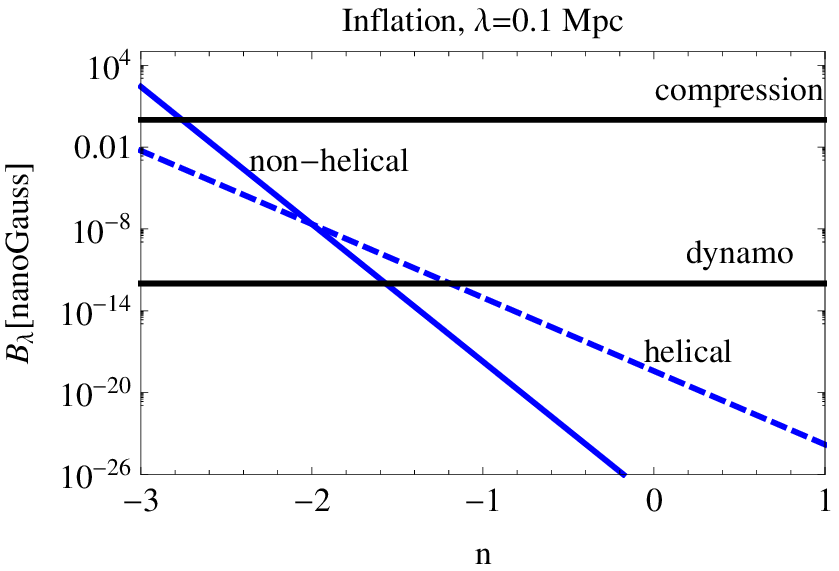} 
\caption{Upper bounds on $B_\lambda$ from GWs and Nucleosynthesis, for non-helical (solid lines) and helical (dashed lines) magnetic fields. Left plot: for a magnetic field generated at a phase transition (PT), with causal spectral index $n=2$, as a function of $\lambda$; red curves are for the QCDPT, and blue curves for the EWPT. Right plot: for a magnetic field generated at inflation, as a function of the spectral index $n$, for $\lambda=0.1$ Mpc. In the helical case, only the part of the plot with $n>-2$ is realistic, because there is no model for the inverse cascade for $n<-2$ \cite{elisa} (note that for an inflationary helical magnetic field in general $n=-1$ \cite{lukas}). The solid, black lines indicate the minimal magnetic amplitude required to seed the fields observed today by compression and by galactic dynamo.} 
\label{Fig1} 
\end{figure}

The condition $\Om_B^*\leq \sqrt{0.1/\mathcal{E}}\,\Om_{\rm rad}$ can be converted into a constraint on $B_\lambda$ (\cf Eq.~\ref{OmB}), which for non-helical fields becomes 
$B_\lambda \lesssim [ 0.1/\mathcal{E}]^{1/4} \,\, \sqrt{\rho_c \, \Om_{\rm rad}} \,/ \, [k_D^* \, \lambda]^{(n+3)/2}~$ \cite{gwold,elisa}. In the case of helical fields, the inverse cascade weakens the bound by a factor $[\eta_{\rm fin}/\eta_L^*]^{(n+2)/3}$, where $\eta_{\rm fin}$ is the final time of the cascade, and $\eta_L^*$ the initial eddy turnover time on the scale of energy injection $L$ \cite{elisa}. Note that the bound on $B_\lambda$ depends on the modeling of the MHD turbulent cascade: in the present analysis, based on \cite{elisa}, a particular model for the cascade has been assumed (\cf section~\ref{sec:model}), and applied to a maximally helical field. The results for $B_\lambda$ are shown in Fig.~\ref{Fig1}: we see that the bound on helical magnetic fields is generically less stringent because of the inverse cascade. Moreover, the dependence on $n$ is such that magnetic fields with blue spectra are more constrained\footnote{Much stronger constraints on $B_\lambda$ can be derived imposing that the magnetic energy cannot overcome 10\% of the total energy in radiation at generation time: $\Om_B^*\leq 0.1 \Om_{\rm rad}$. However, accounting for GW production seems to us more model independent: since GWs do not interact with the rest of the universe, GW energy sourced before Nucleosynthesis is certainly present at Nucleosynthesis time. We can therefore apply to GWs the Nucleosynthesis bound regardless of the epoch of generation of the magnetic field, whether it is at inflation or during the radiation dominated universe \cite{elisa}.}. 

\section{Conclusions}

We have briefly revised how to model a primordial magnetic field and its time evolution. Two properties can be inferred from this simple description. First, causal generation mechanisms give rise to blue spectra, which suppresses the magnetic field amplitude at large scales: they are therefore disfavoured as seeds for the large scale fields observed today in structures. Second, helical fields might circumvent this problem because they undergo an inverse cascade, transferring power to large scales: they are therefore in principle preferred to play the role of seed fields. 

Moreover, if a magnetic field is present in the early universe, it is possible to use observables such as the CMB or Nucleosynthesis to infer limits on its intensity as a function of the parameters describing the field. Recent analyses of the effect of a primordial magnetic field on the CMB show that present CMB data constrain the amplitude of a (generically produced) primordial magnetic field to be less than a few nG, and seem to prefer negative magnetic spectral indexes. At the same time, GWs at Nucleosynthesis can also be used to infer strong constrains on the magnetic field amplitude, and in turns on the proposed generation mechanisms occurring before Nucleosynthesis. It results that a magnetic field generated at inflation with a very red spectrum is poorly constrained, and could be high enough to explain present observations by adiabatic compression during structure formation. Conversely, a magnetic field generated at the EW or QCD phase transitions is much more constrained, and 
by no means it can explain present observations by simple adiabatic compression. It could possibly act as a seed for the galactic dynamo if it is helical, and provided that a smoothing scale of a few kpc is enough to seed the dynamo. In conclusion, in order to explain the magnetic fields observed today, limits for primordial magnetic fields from the CMB and from GWs at Nucleosynthesis point towards a non-causal generation mechanism, occurring at inflation and giving rise to a magnetic field with red spectrum.

\end{document}